\documentstyle[emulateapj]{article}

\begin{document}
\title{Low Luminosity Gamma-Ray Bursts as a Unique Population:
Luminosity Function, Local Rate, and Beaming Factor}
\author{
Enwei Liang$^{1,2}$, Bing Zhang$^1$, Francisco Virgili$^1$,  Z. G. Dai
$^{3,1}$}

\affil{$^1$Department of Physics and Astronomy, University of Nevada, Las Vegas, NV
89154, USA; lew@physics.unlv.edu,bzhang@physics.unlv.edu\\
$^2$Department of Physics, Guangxi University, Nanning 530004, China\\
$^3$Department of Astronomy, Nanjing University, Nanjing 210093,
China\\}

\begin{abstract}
{\em Swift}/BAT has detected $\sim 200$ long-duration GRBs, with redshift measurements
for $\sim 50$ of them. We derive the luminosity function ($\Phi^{\rm HL}$) and the local
event rate ($\rho_0^{\rm HL}$) of the conventional high luminosity (HL) GRBs by using the
$z$-known {\em Swift} GRBs. Our results are generally consistent with that derived from
the {\em CGRO}/BATSE data. However, the fact that {\em Swift} detected a low luminosity
(LL) GRB, GRB 060218, at $z=0.033$ within $\sim 2$ year of operation, together with the
previous detection of the nearby GRB 980425, suggests a much higher local rate for these
LL-GRBs. We explore the possibility that LL-GRBs as a distinct GRB population from the
HL-GRBs. We find that $\rho_0^{\rm LL}$ is $\sim 325_{-177}^{+352}$ Gpc$^{-3}$ yr$^{-1}$,
which is much higher than $\rho_0^{\rm HL}$($1.12_{-0.20}^{+0.43}$ Gpc$^{-3}$ yr$^{-1}$).
This rate is $\sim 0.7\%$ of the local Type Ib/c SNe. Our results, together with the
finding that less than 10\% of Type Ib/c SNe are associated with off-beam GRBs, suggest
that LL-GRBs have a beaming factor typically less than 14, or a jet angle typically wider
than 31$^{\rm o}$. The high local GRB rate, the small beaming factor, and low luminosity
make the LL-GRBs distinct from the HL-GRBs. Although the current data could not fully
rule out the possibility that both HL- and LL-GRBs are the same population, our results
suggest that LL-GRBs are likely a unique GRB population and the observed low redshift GRB
sample is dominated by the LL-GRBs.
\end{abstract}



\section{Introduction}
Gamma-ray bursts (GRBs) and supernovae (SNe) are two of the most violent
explosions in the Universe.  The connection between long duration GRBs and SNe
was predicted theoretically (Colgate 1974; Woosley 1993), and has been verified
observationally through detecting spectroscopic features of the underlying SNe in
GRB 980425/SN 1998bw (Galama et al. 1998; Kulkarni et al. 1998), GRB 030329/SN
2003dh (Stanek et al. 2003; Hjorth et al. 2003), GRB031203/SN 2003lw (Malesani et
al. 2004), and GRB 060218/SN 2006aj (Modjaz et al. 2006; Pian et al. 2006;
Sollerman et al. 2006; Mirabal et al. 2006; Cobb et al. 2006). In some other
cases, red SNe bumps were observed in the late optical afterglow light curves
(Bloom et al. 1999, 2002; Della Valle et al. 2003; Fynbo 2004; see a
comprehensive sample and references in Zeh et al. 2004). It is long speculated
that long GRBs are associated with deaths of massive stars, and hence, SNe (for
recent reviews, see Zhang \& M\'esz\'aros 2004; Piran 2005; M\'esz\'aros 2006;
Woosley \& Bloom 2006). This speculation was broken down by the recent
observation of GRB 060614, which is a long, nearby GRB without an accompanied SN
(Gehrels et al. 2006, Fynbo et. al. 2006, Della Valle et al. 2006, Gal-Yam et al.
2006). Based on its close analogy with the short duration GRB 050724, Zhang et
al. (2007a) argued that this event might be also powered by a merger of compact
stars. They further suggest that the conventional long {\em vs.} short GRB
classification scheme may be modified as Type I (from mergers) {\em vs.} Type II
(from collapsars) GRBs, and that GRB 060614 belongs to Type I (see also Zhang
2006).

Some authors have attempted to determine the luminosity function ($\Phi$) and the local
rate ($\rho_0$) of long GRBs through fitting the $\log N-\log P$ or $V/V_{\rm max}$
distributions observed by CGRO/BATSE (Schmidt 2001; Stern et al. 2002; Lloyd-Ronning et
al. 2002; Norris 2002; Guetta et al. 2005) or through simulations (Lloyd-Ronning et al.
2004; Dai \& Zhang 2005; Daigne et al. 2006). They consider only high luminosity GRBs
(HL-GRBs, with luminosity $L>10^{49}$ erg s $^{-1}$), and generally characterize $\Phi$
with a broken power law and obtain $\rho_0^{\rm HL}\sim 1 ~{\rm Gpc^{-3}~yr^{-1}}$
(e.g. Schmidt 2001; Guetta et al. 2004, 2005). Guetta et al. (2004) suggested that
by extrapolating the LF of HL GRBs to low-luminosities with a broken power law, one gets
$\rho_0^{\rm LL}$ as $\sim 10$ Gpc$^{-3}$ yr$^{-1}$ for low redshift GRBs at
$z<0.17$ (GRBs 980425, 031203, and 030329). However, the fact that {\em Swift} detected
GRB(XRF) 060218 at $z=0.033$ (Mirabal et al 2006) within $\sim 2$ years of operation,
together with the previous discovery of GRB 980425 at $z=0.0085$ (Tinney et al. 1998) by
BeppoSAX, suggests that the local rate of the GRB 060218-like low luminosity (LL) GRBs
($\rho_0^{LL}$) is $100 \sim 1000$ Gpc$^{-3}$ yr$^{-1}$ (Cobb et al. 2006; Pian et
al. 2006; Soderberg et al. 2006b), much higher than $\rho_0^{\rm HL}$ (Schmidt
2001; Stern et al. 2002; Lloyd-Ronning et al. 2002; Norris 2002; Guetta et al. 2004,
2005). This poses a great puzzle about the nature of of these LL-GRBs. Are these LL-GRBs
from the same population as HL-GRBs (Nakamura 1999; Ioka \& Nakamura 2001; Yamazaki et
al. 2003; Guetta et al. 2004) or from a sub-energetic GRB population (Kulkarni et al.
1998; Soderber 2004b; Mazzali et al. 2006; Toma et al. 2006) with a much higher rate than
that of the conventional HL-GRBs? In this paper we attempt to address these questions
through careful statistical analyses with the redshift-known {\em Swift} GRB sample.
Throughout the paper $H_0=71$ km s$^{-1}$ Mpc$^{-1}$, $\Omega_{\rm{m}}=0.3$, and
$\Omega_\Lambda=0.7$ are adopted.

\section{Data and GRB sample}
The current GRB sample with redshift measurements contains more than 80 GRBs. Since both
the observed luminosity and redshift distributions are instrument-dependent (Jakobsson et
al. 2006; Zhang et al. 2007b), we use only the long GRBs detected by {\em Swift} to form a
homogeneous sample. {\em Swift} has detected $\sim 200$ long bursts during the first 2
years of operation (from December 2004 to November 2006). Among them 47 bursts have
redshift measurements. We have excluded GRBs 060505 and 060614 from our sample since they
are suggested to be classified as a new type of GRBs (Gehrels et al. 2006; Gal-Yam et al.
2006; Fynbo et al. 2006; Della Valle et al. 2006) or to belong to the short GRB group
(Zhang et al. 2007a; Zhang 2006). We therefore obtain a {\em Swift} HL-GRB sample of 45 bursts.
We collect the peak flux and spectral parameters of these GRBs from the literature or GCN
Circular reports. It is well known that most broad band GRB spectra are well fitted by
the Band function (Band et al. 1993). However, $\sim 80\%$ of the BAT spectra are fitted
by a simple power law ($F_\nu \propto \nu^{-\Gamma^{PL}}$) (e.g., Zhang et al. 2007b).
This effect is due the narrowness of the BAT band (15-150 keV) and the faintness of the
bursts (so that there are not enough counts to perform a Band-spectrum fit. Therefore,
the BAT spectra are not the true spectra of the {\em Swift} GRBs. In order to correct the
observed luminosity to a broad band ($1-10^4$ keV in this analysis) we have to estimate
the parameters of the Band function for each GRB. Zhang et al. (2007b) developed a new
method to derive the parameters of the Band function by fitting the observational data
with incorporation of the observed spectral hardness ratio. They found a strong
correlation between $\Gamma^{PL}$ derived from the BAT data and $E_p$ of the $\nu f_\nu$
spectrum (see also Zhang et al. 2007a). Sakamoto et al. (2006) independently derived this
correlation. We use this method to estimate the parameters of the Band function for the
GRBs in our sample, and then correct the observed luminosity to $1-10^4$ keV band in the
burst rest frame with the $k$-correction presented by Bloom et al. (2001). The HL-GRB
sample has a moderate size, with both a broad luminosity distribution ($L_{iso}= 2\times
10^{50}\sim 4.6\times 10^{53} ~{\rm erg~s^{-1}}$) and a broad redshift distribution
($z=0.55\sim 6.29$).

\section{Analysis Method}
The number of GRBs per unit time at redshift $z\sim z+dz$ with
luminosity $L\sim L+dL$ is given by
\begin{equation}\label{dn}
\frac{d N}{dtdzdL}=\frac{R_{\rm GRB}(z)}{1+z}\frac{dV(z)}{dz}\Phi(L),
\end{equation}
where $R_{\rm GRB}(z)$ is the event rate (in units of ${\rm Gpc^{-3} yr^{-1}}$) as a
function of $z$, $(1+z)^{-1}$ accounts for the cosmological time dilation, and
\begin{equation}\label{dvdz}
\frac{dV}{dz}=\frac{c}{H_0}\frac{4\pi D^2_L}{(1+z)^2
[\Omega_M(1+z)^3+\Omega_\Lambda]^{1/2}}
\end{equation}
is the comoving volume element at redshift $z$ in a flat $\Lambda$ CDM
universe. We assume that $R_{\rm GRB}$ follows the star formation rate
as a function of redshift, and the parameterized star formation model
SF2 presented by Porciani \& Madau (2001) is used\footnote{We have
considered the other two star formation models, SF1 and SF3. Our
results are not sensitive to the selection of these star formation
models since these models are almost the same at low redshifts.},
\begin{equation}\label{SF2}
R_{\rm GRB}={23\rho_0}\frac{\rm {e}^{3.4z}}{\rm{e}^{3.4z}+22.0},
\end{equation}
where $\rho_0$ is the underlying local GRB rate per unit volume at $z\sim 0$, i.e.,
$\rho_0=R_{\rm GRB}|_{z\sim 0}$. We characterize $\Phi(L)$ as
\begin{equation}\label{LF}
\Phi(L)=\Phi_0\left[\left(\frac{L}{L_b}\right)^{\alpha_1}
+\left(\frac{L}{L_b}\right)^{\alpha_2}\right]^{-1},
\end{equation}
where $L_b$ is the break luminosity and $\Phi_0$ is a normalization
constant to assure the integral over the luminosity function being
equal to unity. Considering an instrument having a flux threshold
$F_{\rm th}$ and an average solid angle $\Omega$ for the aperture
flux, the number of the detected GRBs after an observational period of
$T$ should be
\begin{equation}\label{N}
N=\frac{\Omega T}{4\pi}\int_{L_{\min}}^{L_{\max}}
\Phi(L)dL\int_0^{z_{\max}} \frac{R_{\rm
GRB}(z)}{1+z}\frac{dV(z)}{dz}dz,
\end{equation}
where $L_{\rm max}$ and $L_{\rm min}$ are taken as $10^{54}$ and $10^{45}$ erg s$^{-1}$,
respectively, and $z_{\max}$ for a given burst with luminosity $L$ is determined by the
instrumental flux threshold $F_{\rm th}$ through $F_{\rm th}=L/4\pi D^2_L(z_{\max})$.

We derive $\Phi(L; \alpha_1, \alpha_2, L_b)$ with the {\em Swift} GRB sample by using the
following procedure. First, we constrain $\alpha_1$, $\alpha_2$, and $L_b$ by comparing
the luminosity and redshift distributions predicted by the model to the observations (1-D
Criterion). We calculate the $L$ and $z$ distributions for a given set of the parameters,
and then measure the consistency between model predictions and observations with the K-S
test (Press et al. 1999). The significance level of the consistency is described by a
probability of $p_{K-S}$. A larger value of $p_{K-S}$ suggests a better consistency (the
higher significance level). The consistency for both $L$ and $z$ distributions is
therefore measured by $p_{K-S}^{T}=p_{K-S}^{L}\times p_{K-S}^{z}$. Second, we derive
$\rho_0$ for a given $\Phi$ by using $N=N^{obs}$ ($N$-Criterion). In this analysis we
consider only the GRBs detected by Swift/BAT. The aperture solid angle of BAT is $1.33$,
and the detection numbers of both HL- and LL-GRBs by Swift/BAT in 2 operation years is
$N^{obs}_{HL}=200$ and $N^{obs}_{\rm LL}=1$. The sensitivity curve of BAT in the 50-150
keV band (Band 2003) is adopted. Third, we use a 2-dimensional GRB number distributions
in the [$\log L, \ \log (z)$]-plane to further examine the parameters and evaluate the
influence of the observational biases of the sample (2-D Criterion). The 1-D Criterion
constrains the parameters with the {\em clustered} bursts in the sample. The 2-D
Criterion, instead, present a self-consistency test with the scattered data points by
assuming that the burst distribution in the [$\log L, \ \log (z)$]-plane predicted by a
reasonable $\Phi$ should cover the observational data points at a $3\sigma$ confidence
level.

\section{Are HL-GRBs and LL-GRBs the Same Population?}
We first study the constraints on $\Phi^{\rm HL}$ with the the current $z-$known HL-GRB
sample, and then analyze whether or not it can predict the detection rate of LL-GRBs by
extrapolating $\Phi^{\rm HL}$ down to a luminosity as that of GRBs 980425 and 060218.

\subsection{Luminosity Function and Local Rate of HL-GRBs}
\begin{figure}
\plotone{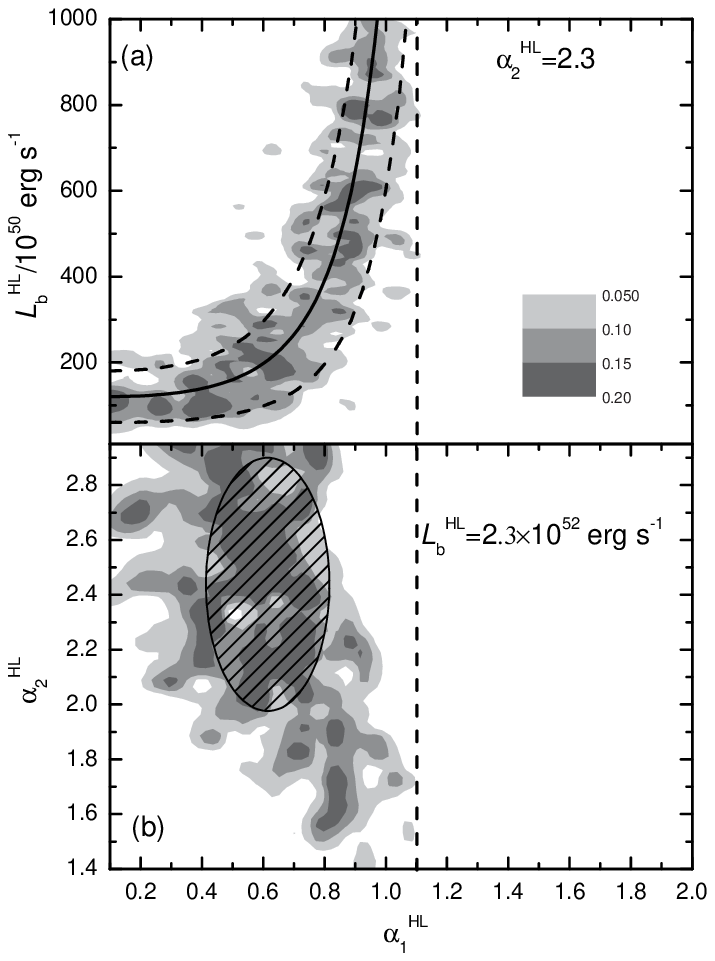} \caption{Distributions of $p_{K-S}^{T}$ in the planes of ($\alpha_1^{\rm
HL}$, $L_b^{\rm HL}$)$|_{\alpha_2^{\rm HL}=2.3}$ (panel a) and ($\alpha_1^{\rm HL}$,
$\alpha_2^{\rm HL}$)$|_{L_b^{\rm HL}=2.3\times 10^{52}}$ (panel b) derived with the
current {\em Swift} HL-GRB sample. The vertical dashed
line marks a limit on $\alpha_1$. The
curves in the panel (a) show the relation and its errors between $L_b^{\rm HL}$ and
$\alpha_1^{\rm HL}$. The ellipse in panel (b) marks the best parameters of $\alpha_1^{\rm
HL}$ and $\alpha_2^{\rm HL}$.
\label{P_ks}}
\end{figure}

Following the methodology discussed above, we derive the luminosity function and
local rate of HL-GRBs with our $z-$known HL-GRB sample. Figure \ref{P_ks}(a) and
(b) show the distribution of $p_{K-S}^{T}$ in ($\alpha_1^{\rm HL}$,
$\alpha_2^{\rm HL}$)$|_{L_b^{\rm HL}=2.3\times 10^{52}}$ and ($\alpha_1^{\rm
HL}$, $L_b^{\rm HL}$)$|_{\alpha_2^{\rm HL}=2.3}$ planes, respectively. Our sample
includes 44 bursts, and a value of $p_{K-S}^{T}>0.05$ marginally suggests the
consistency between the model predictions and the observations. Therefore, we
show the contours for $p_{K-S}^{T}>0.05$ only. From Fig.\ref{P_ks} we find that
the parameters are convolved. Different combinations of these parameters could
make the consistency between both the luminosity and redshift distributions and
the observations. Nonetheless, strong constraints on the parameters could be
derived from Fig.\ref{P_ks}. First, the value of $\alpha_1^{\rm HL}$ is required
to be smaller than 1.1 (marked by a vertical line in the two panels). Second,
$L_b^{\rm HL}$ is correlated with $\alpha_1^{\rm HL}$. This correlation is
roughly described by a function of $\log L_b^{\rm HL}/(1.2\pm
0.6)10^{52}={\alpha^{\rm HL}_1}^3$ [marked by solid and dashed curves in Fig.
\ref{P_ks}(a)]. Third, $\alpha_2^{\rm HL}$ should be larger than 1.5, and a
preferred value from our sample is $2.0<\alpha_2^{\rm HL}<2.6$ and
$0.5<\alpha_1^{\rm HL}<0.8$ [with $p_{K-S}^{T}>0.15$, marked by an ellipse in
Fig. \ref{P_ks}(b)].

\begin{figure}
\plotone{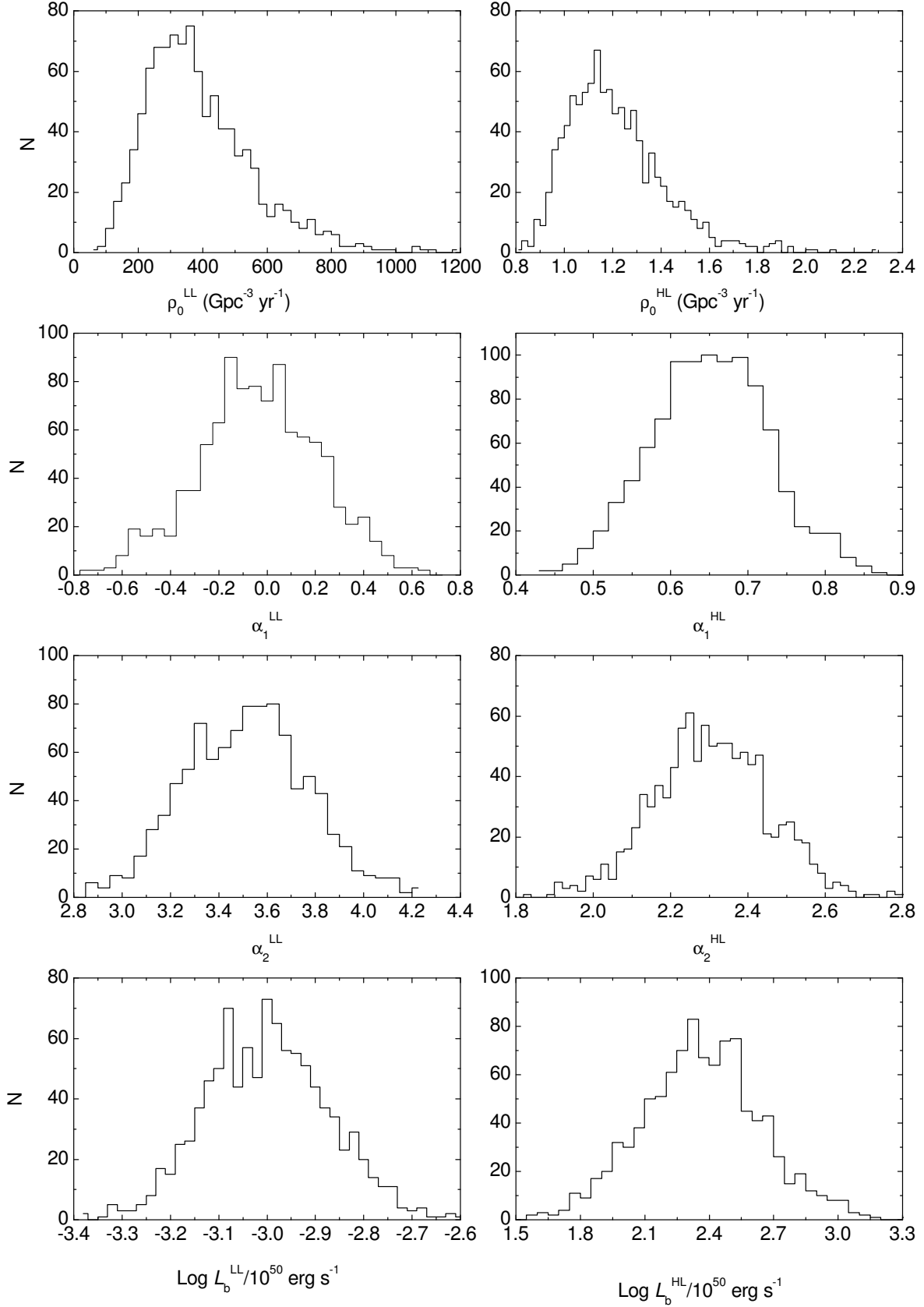}
\caption{Distributions of the GRB local event rates and the parameters of the LFs
for LL-GRBs ({\em left panels}) and HL-GRBs ({\em right panels}).
\label{LF_dis}}
\end{figure}

With these constraints on the parameters we derive the distribution of
$\rho_0^{\rm HL}$ through simulations. We take $\alpha_1^{\rm HL}=0.65\pm 0.15$,
$\alpha_2^{\rm HL}=2.3\pm 0.3$, and $L_b^{\rm HL}/10^{52}=(1.2\pm 0.6)\times
10^{\alpha_1^3}$. We assume a normal distribution of the errors for
$\alpha_1^{\rm HL}$, $\alpha_2^{\rm HL}$, and $\log L_b^{\rm HL}$, and simulate
1000 sets of ($\alpha_1^{\rm HL}$, $\alpha_1^{\rm HL}$, $\log L_b^{\rm HL}$). We
then calculate $\rho_0^{\rm HL}$ for each set of the parameters with Eq. \ref{N}
by taking $N=N^{\rm HL}_{\rm obs}$. The distribution of the $\rho_0^{\rm HL}$
together with that of these parameters are shown in Fig. \ref{LF_dis}. We obtain
$\rho_0^{\rm HL}=1.12_{-0.20}^{+0.43}$ Gpc$^{-3}$ yr$^{-1}$ at a $90\%$
confidence level. The 1-dimensional and 2-dimensional distributions of the
luminosity and redshift predicted by our model with $\alpha_1=0.65$,
$\alpha_2=2.3$,$L_b=2.25\times 10^{52}$ erg s$^{-1}$ are shown in Fig. \ref{2_D}.
It is found that both the one-D $z$ and $L$ distributions predicted by our model
are consistent with the observations, and all the HL-GRBs in the 2-dimensional
plane are enclosed in the $3\sigma$ region. Both $\Phi^{\rm HL}$ and $\rho_0^{\rm
HL}$ are consistent with those derived from the BATSE data (e.g. Schmidt 2001;
Stern et al. 2002; Guetta et al. 2005) and simulations (e.g. Dai \& Zhang 2005;
Daigne et al. 2006).

\subsection{LL-GRBs as the Low Luminosity End of the HL-GRB Population?}
We estimate $\rho_0^{\rm LL}$ with the two detections of GRB 060218 and GRB
980425 from,
\begin{equation}\label{NV}
N=\rho_0^{\rm LL} V_{z<0.033}\frac{\Omega^{Bepp}T^{Bepp} +\Omega^{Sw} T^{Sw}}{4\pi}=2,
\end{equation}
where $V_{z<0.033} \sim 1.2\times 10^{-2}$ Gpc$^{3}$ is the volume at $z<0.033$,
$\Omega^{Bepp}=0.123$ and $\Omega^{Sw}=1.33$ are the solid angles of the GRBM on
board {\em BeppoSAX} and the BAT on board {\em Swift}, respectively, and
$T^{Bepp}\sim 6$ years and $T^{Sw}\sim 2$ years (at the time when this paper is
written) are the operation periods of the {\em BeppoSAX} and {\em Swift},
respectively. We obtain $\rho_0^{\rm LL} \sim 600~{\rm Gpc}^{-3}~{\rm yr}^{-1}$.
This is much larger than $\rho_0^{\rm HL}$ derived above(see also Cobb et al.
2006 and Soderberg et al. 2006b). A simple extrapolation of $\Phi^{\rm HL}$
derived above to low luminosities cannot account for such a high $\rho_0^{\rm
LL}$ (e.g. $\rho_0^{\rm LL} \sim 10~{\rm Gpc}^{-3}~{\rm yr}^{-1}$; Guetta et al.
2004). This rate yields only a Poisson probability of $5.1\times 10^{-4}$ for
detecting the two events at $z<0.0331$. In order to argue a same population for
both HL and LL GRBs and keep a high $\rho_0^{\rm LL}$, the only way is to invoke
a much larger $\alpha_1$ than $\alpha_1^{\rm HL}$ derived above. We fix
$\alpha_2=2.3$, $L_b=5\times 10^{52}$ erg s$^{-1}$, and $N=200$, and search for a
$\alpha_1$ value that can yield a local rate as $\sim 600~{\rm Gpc}^{-3}~{\rm
yr}^{-1}$. We obtain $\alpha_1\sim 1.6$. The derived one-D and two-D
distributions of the redshift and luminosity are shown in Fig.\ref{2_D_one}. One
can see that the model predictions are significantly deviated from the
observations. The model greatly over-predicts GRBs with $z\sim 1.2$ and $L\sim
3\times 10^{52}$ erg s$^{-1}$. Although this result could not fully rule out the
possibility that LL-GRBs are simply the low luminosity end of the HL-GRB
population(Guetta et al. 2004), it does disfavor such a possibility. In the above
analysis we do not consider the difference of the beaming effects for both HL-
and LL-GRBs and the LF cosmological evolution effect. As we discuss below, the
LL-GRBs are likely less collimated and are detectable in the nearby universe
only.

\section{LL-GRBs as a Distinct GRB Population from HL-GRBs}
As discussed above, the high detection rate of the LL-GRBs motivates us to consider the
LL-GRBs as a distinct GRB population from the HL-GRBs. The conventional HL-GRBs generally
have a luminosity of $L>10^{49}$ erg $s^{-1}$. We therefore take a preliminary criterion
of $L<10^{49}$ erg $s^{-1}$ to select our LL-GRB sample. LL-GRBs are faint. They are only
detectable in a small volume of the local universe and a large amount of the population
are below the sensitivity threshold of the detector. The observable LL-GRBs with {\em
Swift} are rare events comparable to HL-GRBs. It is unlikely to establish a large sample
with the current GRB missions, so it is difficult to investigate $\Phi^{\rm LL}$ through
fitting its $\log N-\log P$ distribution or through our 1-D Criteria (as is done for the
HL-population). We can only roughly constrain the $\Phi^{\rm LL}$ and $\rho_0^{\rm LL}$
with a few detections and limits of LL-GRBs. GRBs 980425 and 060218 are two firm
detections of LL-GRBs\footnote{Please note that GRB 060218 shows significant hard-to-soft
spectral evolution (Campana et al. 2006; Ghisellini et al. 2006) and the peak energy of
its integrated spectrum matches the Amati-relation (Amati et al. 2006). GRB 980425
significantly deviates this relation. Ghisellini et al. (2006) argued that by considering
the spectral evolution effect, GRB 980425 may be also consistent with the
Amati-relation.}. There are also two other marginal detections for the LL-GRBs, i.e.,
GRBs 031203 ($z=0.105$, $L=3.5\times 10^{48}$ erg s$^{-1}$) and 020903 ($z=0.25$,
Soderberg et al. 2002; $L=8.3\times 10^{48}$ erg s$^{-1}$).

\subsection{Luminosity Function and local Rate}
With the four detections and the other constraints from observations we constrain the LF
of these LL-GRBs. The luminosity of these LL-GRBs range from $5\times 10^{46}$ erg
s$^{-1}$ to $8.3\times 10^{49}$ erg s$^{-1}$. Assuming also a broken power law LF for the
LL-population (similar to eq.[4]), we take $L_b$ around $10^{47}$ erg s$^{-1}$ and
constrain $\alpha_1$ and $\alpha_2$ by requiring that the $3\sigma$ contour of the 2-D
distribution encloses these LL-GRBs. This places constraints on both $\alpha_1$ and
$\alpha_2$. In order to make the $3\sigma$ contour marginally enclose the nearest burst,
GRB 980425, but not over-predict the detection probability at $z<0.01$, $\alpha_1$ should
be shallow. Similarly, the $\alpha_2$ is constrained by GRBs 031203 and 020903. Based on
these observational constraints we search for $\alpha_1^{\rm LL}$ and $\alpha_2^{\rm LL}$
by taking $L_b^{\rm LL}=(1.0\pm 0.3)\times 10^{47}$ erg s$^{-1}$. We find $\alpha_1^{\rm
LL}=0\pm 0.5$ and $\alpha_2^{\rm LL}\sim 3.0\sim 4.0$ can roughly reflect these
constraints. We use the same simulation method as that for HL-GRBs to derive the
distribution of $\rho_0^{\rm LL}$. The parameters are taken as $\alpha_1^{\rm LL}=0\pm
0.5$, $\alpha_2^{\rm LL}=3.5\pm 0.5$, and $L_b^{\rm LL}=(1.0\pm 0.3)\times 10^{47}$ erg
s$^{-1}$. The distribution of $\rho_0^{\rm LL}$ together with that of these parameters
are also shown in Fig. \ref{LF_dis}. We obtain $\rho_0^{\rm LL}=325_{-177}^{+352}$ at a
$90\%$ confidence level. The 2-D distribution in the $(\log L,\log z)$ plane is shown
Fig. \ref{2_D}. It is found that the LL-GRBs form a distinct ``island'' from the main
``continental'' population. The detection rate of the LL-GRBs thus can be explained
without over-predicting the HL-GRBs. These results suggest that the current data are
consistent with the conjecture that LL-GRBs form a distinct population from HL-GRBs, with
a low luminosity and a high local rate. The constrained luminosity functions for both HL
and LL populations are displayed in Fig.\ref{Rate_obs}(a).

\subsection{Beaming factor}
Understanding GRB rates has profound implications for understanding the relation
between GRBs and Type Ib/c SNe (e.g. Lamb et al. 2005). Compared with the local
Ib/c SN rate ($4.8\times 10^4 {\rm Gpc}^{-3} \rm{yr}^{-1}$; Marzke et al. 1998;
Cappellaro, Evans \& Turatto 1999; Folkes et al. 1999), the rate of LL-GRBs
(on-beam only, not including those beamed away from the Earth) is about $\sim
0.7\%$ of the Type Ib/c SN rate. Most recently, Soderberg et al. (2006a) argued
that at most $\sim 10\%$ of Type Ib/c SNe are associated with off-beam LL-GRBs
based on their late-time radio observations of 68 local Type Ib/c SNe. This
result, combined with our result, suggest that the beaming factor of these
LL-GRBs is at most a factor of 14, in contrast to a higher factor ($\sim 100$,
Frail et al. 2001; Zhang et al. 2004; Guetta et al. 2005) for HL-GRBs. This
suggests that LL-GRBs are less collimated, with an opening angle typically larger
than $\sim 31^\circ$. This is consistent with the observational data of GRB
060218 (Campana et al. 2006; Soderberg et al. 2006b).

\section{Observed local GRB event rates of HL-GRBs and LL-GRBs}
\begin{figure}
\plotone{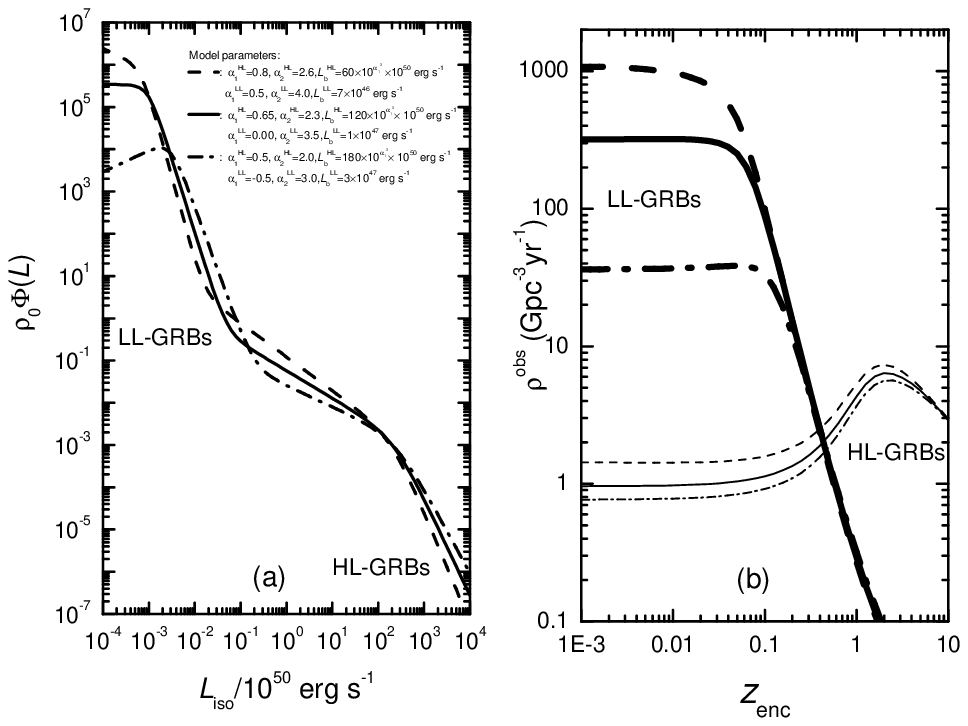} \caption{{\em Panel (a):} The combined LFs of both LL- and HL- GRBs
derived from a set of {\em ordinary} parameters (solid line)  and from two sets of
parameters that are roughly regarded as the lower (dash-dotted line) and upper (dashed
line) limits of the LFs. {\em Panel (b):} The observed GRB event rates for both LL- and
HL-GRBs as a function of ``enclosing redshift'' $z_{enc}$ (i.e. the volume enclosed by
this redshift) for the three parameter sets shown in panel (a). Same line styles for
different models are adopted in both panels.
 \label{Rate_obs}}
\end{figure}

The so-called {\em observed} local GRB event rate $\rho^{\rm obs}$ crucially
depends on the luminosity function, the instrument threshold, and a certain
redshift to estimate the rate. Figure \ref{Rate_obs}(a) shows the combined $\Phi$
of both LL- and HL- GRBs derived from a set of {\em ordinary} parameters
($\alpha_1^{\rm HL}=0.65$, $\alpha_2^{\rm HL}=2.3$, $L_b=1.20\times 10^{52}
\times 10^{{\alpha_1^{\rm HL}}^3}$ erg s$^{-1}$; $\alpha_1^{\rm LL}=0.0$,
$\alpha_2^{\rm LL}=3.5$, $L_b^{\rm LL}=10^{47}$ erg s$^{-1}$) and from two sets
of parameters that are roughly taken as the lower limit ($\alpha_1^{\rm HL}=0.5$,
$\alpha_2^{\rm HL}=2.0$, $L_b^{\rm HL}=0.6\times 10^{52} \times
10^{{\alpha_1^{\rm HL}}^3}$ erg s$^{-1}$; $\alpha_1^{\rm LL}=-0.5$,
$\alpha_2^{\rm LL}=4.0$, $L_b^{\rm LL}=7 \times 10^{46}$ erg s$^{-1}$) and upper
limit ($\alpha_1^{\rm HL}=0.8$, $\alpha_2^{\rm HL}=2.6$, $L_b^{\rm HL}=1.8\times
10^{52} \times 10^{{\alpha_1^{\rm HL}}^3}$ erg s$^{-1}$; $\alpha_1^{\rm LL}=0.5$,
$\alpha_2^{\rm LL}=3.0$, $L_b^{\rm LL}=3 \times 10^{47}$ erg s$^{-1}$) of the LF.
It is evident that $\Phi$ has two distinct components. We calculate the observed
GRB event rates for both LL- and HL-GRBs as a function of ``enclosing redshift''
$z_{enc}$ (i.e. the volume enclosed by this redshift) for the three parameter
sets. The results are shown in Fig.\ref{Rate_obs}(b). The observed rates of
LL-GRBs and HL-GRBs as a function of redshift are significantly different.
$\rho_{\rm obs}^{\rm LL}$ keeps almost a constant within $z<0.1$, and drops
sharply at $z>0.1$. At $z\sim 0.2$ $\rho_{\rm obs}^{\rm LL}$ is $\sim 10$
Gpc$^{-3}$ yr$^{-1}$, which is roughly consistent with that suggested by Guetta
et al. (2004). The $\rho_{\rm obs}^{\rm HL}$ is also almost constant within
$z<0.1$, but it increases when $z>0.1$ and peaks at $z\sim 2.5$, being consistent
with the observations with {\em Swift}. Although $\rho_{obs}^{\rm LL}(z)$ has a
large uncertainty, (e.g. varies by almost two orders of magnitude),
Fig.\ref{Rate_obs}(b) indicates that the detectable GRB sample within $z<0.4$ is
dominated by the LL-GRBs.

\section{Discussion}
We have analyzed the luminosity function and local rate of HL-GRBs and extended this
analysis to LL-GRBs. We present detailed discussion on our method, data, and analysis
results before drawing the conclusions of our analysis.

\begin{figure}
\plotone{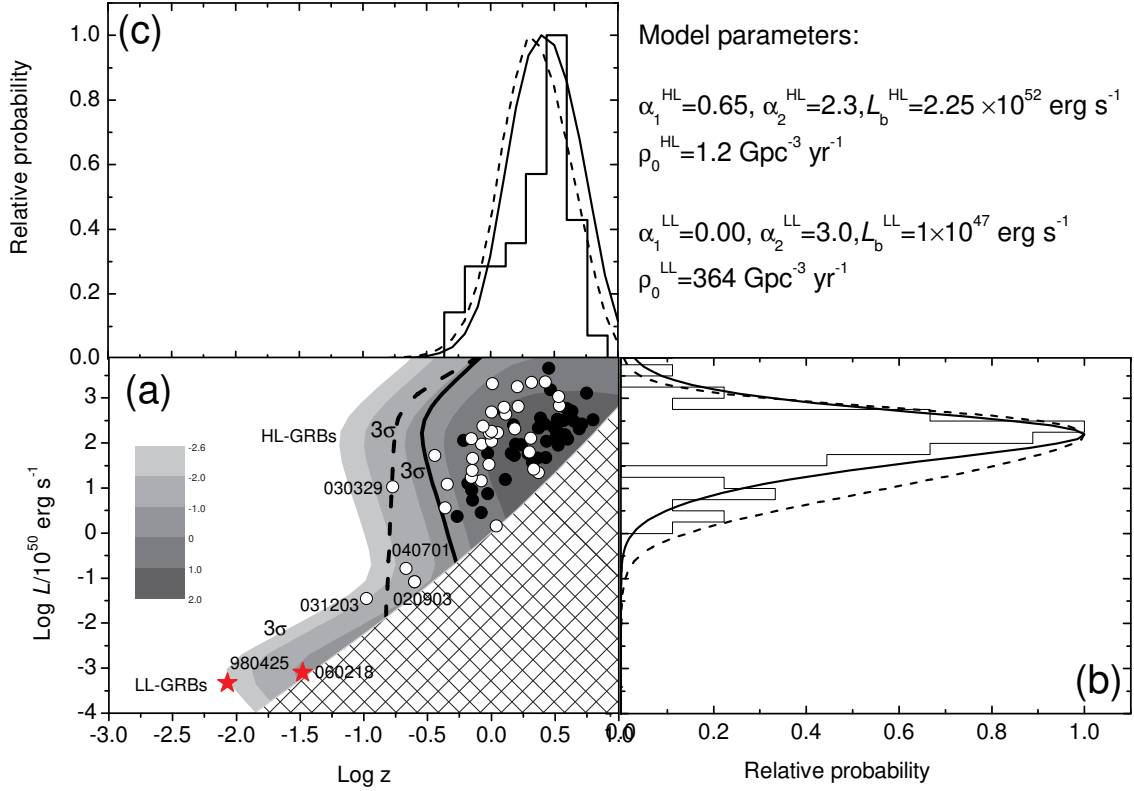} \caption{Jointed contours of the logarithmic GRB detection rate [$\log
(dN/dt)$] distribution in a 2-dimensional (2-d) [$\log L$, $\log (z)$]-plane as compared
with observational data [panel (a)], assuming that the HL- and LL- GRBs are two distinct
populations. The two firm LL-GRBs are denoted by stars, and the {\em Swift} HL-GRBs are
denoted by solid dots. The hatched region marks the limitation of the {\em Swift}/BAT
detectability, where the threshold is derived by using the {\em Swift}/BAT sensitivity in
the 50-150 keV band for a {\it standard} GRB with $E_p=200$ keV in the GRB local frame.
The bold-faced, solid curves in panel (a) marks the $3\sigma$ confidence level of the 2-d
distributions for the HL- and LL-GRBs. The comparisons of the observed 1-d distributions
of $\log L$ and $\log z$ with the model predictions are presented in panels (b) and (c),
respectively. The dashed curve in the panel (a) and the dashed lines in the panels (b)
and (c) are, respectively, the $3\sigma$ contour of the 2-d distribution and the
corresponding 1-d distributions derived from a LF with $\alpha_1^{\rm HL}=1.05$,
$\alpha_2^{\rm HL}=3$, and $\L_b^{\rm HL}=6\times 10^{52}$ erg s$^{-1}$, which gives a
$3\sigma$ contour that can enclose all the HL-GRBs observed by {\em Swift} and pre-{\em
Swift} missions ({\em see \S 7 in the text}). \label{2_D}}
\end{figure}

\begin{figure}
\plotone{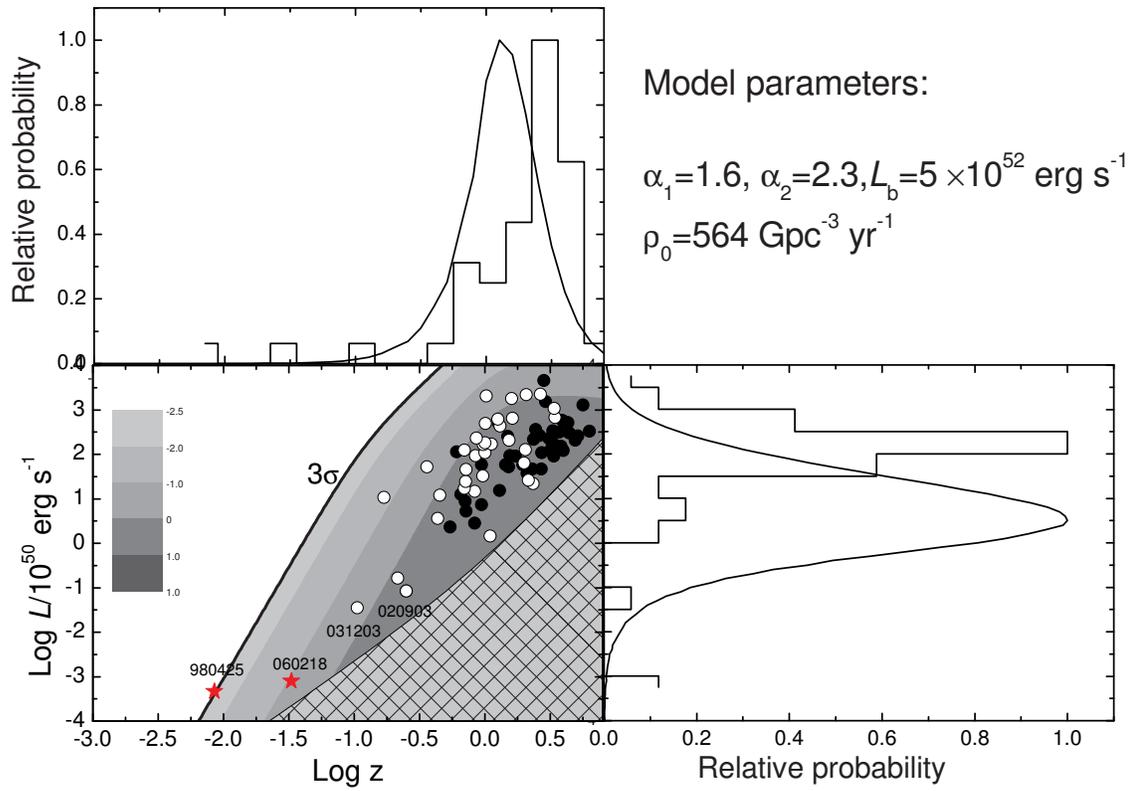} \caption{Same as Fig.\ref{2_D} but for the case that the HL- and LL- GRBs
are assumed to belong to the same population.
 \label{2_D_one}}
\end{figure}

Ideally, deriving the luminosity function requires a large, unbiased $z-$known sample.
The current GRB sample, however, is still small and suffers from some observational
biases. First, the trigger probability of a burst depends on its peak flux. A weaker
burst near the instrument threshold has a lower trigger probability. This effect affects
the completeness of the sample at a given instrumental threshold. As shown in
Fig.\ref{2_D}, a gap between the data points and the threshold curve is observed. The gap
manifests this bias. Second, the redshift measurements also have biases against weak GRBs
(e.g. Bloom 2003). In order to evaluate whether these observational biases could
significantly affect our results, we use the 2-D distribution in the $(\log L,\log z)$
plane at 3$\sigma$ significance level to examine our results. If these biases do not
ruinously affect our results, the 3$\sigma$ contour of the GRB distribution in the $(\log
L,\log z)$ plane should enclose the current GRB sample; otherwise, the model should
predict many undetected GRBs near the threshold and the 3$\sigma$ contour cannot cover
the scattered data point in the $(\log L,\log z)$ plane. From Fig.\ref{2_D} we find the
data points are enclosed by the $3\sigma$ contour for the HL-GRBs, suggesting that these
observational biases do not significantly affect our results for HL-GRBs. However, the
current {\em Swift} GRB sample with redshift measurements is still small. One prominent
issue is whether both the redshift and luminosity distributions derived from the current
{\em Swift} GRB sample represent the true {\em observed} ones for the threshold of {\em
Swift}. In order to address this issue we plot also the GRBs detected by previous GRB
missions in Fig.\ref{2_D}. It is found that most of these GRBs are enclosed by the
3$\sigma$ contour defined with the current {\em Swift} HL-GRB sample. Two GRBs, 030329
and 040701, are out of the $3\sigma$ contour. In order to make a $3\sigma$ contour to
enclose the two GRBs, we find that the luminosity function should have $\alpha_1>1.05$,
$\alpha_2\sim 3.0$, and $L_b\sim 6\times 10^{52}$ erg s$^{-1}$. We also show that the
$3\sigma$ contour and the predicted $L$ and $z$ distributions of this luminosity function
in Fig.\ref{2_D} (dashed line). It predicts more GRBs with $L\sim 10^{49}-10^{51}$ erg
s$^{-1}$ (roughly the low luminosity end of the HL-GRBs) than observations by {\em
Swift}. If this is indeed an observational bias for the HL-GRB sample with {\em Swift},
the true $\alpha_1$ should be larger than that derived from the current {\em Swift} GRB
sample, say, $\alpha_1\sim 1$. With such a LF, GRB 020903 is enclosed in the $3\sigma$
contour and 031203 also marginally belongs to the HL-GRB group. However, even with such a
LF, GRBs 980425 and 060218 are definitely not merged into the HL-GRB group.

Another effective approach to evaluate the biases is to fit the observed $\log N-\log P$
distribution. The current {\em Swift} sample has only $\sim 200$ bursts. It still cannot
be used for making such a fitting. A sample of $\sim 500$ bursts with $\sim 120$ redshift
measurements could be obtained by {\rm Swift} in a period of 5 years. Such a sample could
refine the constraints on the LF of HL-GRBs by incorporating our approach with the $\log
N-\log P$ fitting\footnote{The $\log N-\log P$ distribution is sensitive to the
instrument threshold. One has to use a sample observed by a given instrument to make such
a fit. {\em CGRO}/BATSE established a large sample and such a fit gives $\alpha_1=0.3\sim
1.0$ and $\alpha_2\sim 2\sim 3$ by different authors. This is consistent with our results
shown in Fig.\ref{P_ks}}.

A cosmological evolution effect may influence the derivation of the luminosity function
(e.g. Lloy-Ronning et al. 2002). So far there is no robust evidence of such an
evolutionary effect with the GRB sample with spectroscopic redshifts. This effect also
does not affect $\Phi^{\rm LL}$, since the LL-GRBs are likely detectable only at low
redshifts. It may influence the derivation of $\Phi^{\rm HL}$. A large sample of
spectroscopic redshift measurements is required to more robustly evaluate this effect.
The current HL-GRB sample is inadequate for this task. Therefore, we do not consider this
effect in this analysis. In the future, using a sample established by {\em Swift} within
5 year operation one may be able to robustly model this effect, and the constraints on
the luminosity function of HL-GRBs could be further refined.

Our results tentatively suggest that LL-GRBs are an intrinsically different GRB
population, characterized by a high local rate, low luminosity, and small beaming
factor comparing to the HL-GRBs. The detectable low redshift GRBs should be
dominated by the LL-GRBs. Le \& Dermer (2007) draw the similar conclusion through
modeling the redshift distribution of the {\em Swift} GRBs. However, we pointed
out that the current data do not conclusively exclude the possibility that only
one population of GRBs is responsible for both HL- and LL-GRBs(Guetta et al.
2004).

There are two scenarios to explain the nature of the LL-GRBs. One scenario is that these
GRBs are standard HL-GRBs viewed off-axially (e.g. Nakamura et al. 1999; Yamazaki et al.
2003). In order to account for the step-like two-component LF displayed in Fig. \ref{LF},
the jet must include two distinct components, i.e. a narrow HL component and a very wide
LL component. Such a jet configuration is different from the conventional jet-cocoon
picture in the standard collapsar model in which the cocoon component is not as broad as
31$^{\rm o}$ (e.g. Zhang et al. 2003).  The lack of detection of radio re-brightening in
GRB 980425 and GRB 031203 (Soderberg et al. 2004a,b) has also greatly constrained this
scenario (Waxman 2004a,b).

The second scenario is that LL-GRBs are intrinsically different from HL-GRBs (e.g. Wang
et al. 2000; Soderberg et al. 2004b; Mazzali et al. 2006). Actually, the LL-GRBs as a
unique GRB population has been proposed since the discovery of GRB 980425/SN1998bw with
its observed under-luminous, long-lag, simple light curve features (e.g., Soderberg et
al. 2004b). Stern et al. (1999) suggested a group of ``simple'' bursts with peak fluxes
near the BATSE trigger threshold. Norris (2002) found that the proportion of long-lag,
single-pulse bursts (similar to the three LL-GRBs) within long-duration bursts increases
from negligible among bright BATSE bursts to$\sim 50\%$ at the trigger threshold, and
their peak flux $\sim 2$ orders of magnitude lower than that of the brightest bursts.
Taken together with the fact that three nearby GRBs, 980425, 031203, and 060218, are
long-lag and under-luminous (e.g., Sazonov et al. 2004; Liang et al. 2007), an intuitive
speculation of the observed LL-GRBs is that that they are probably relatively nearby
(e.g., Norris et al. 2005) and their local rate should be much higher that conventional
HL-GRBs as we derived in this analysis. This scenario calls for a different type of
progenitors (e.g. neutron stars rather than black holes) for LL-GRBs from those of
HL-GRBs (e.g. Mazzali et al. 2006; Soderberg et al. 2006b). The abundant multi-wavelength
observations of GRB 060218 lead us make a step toward to the nature of this kind of
events (Wang \& M\'{e}sz\'{a}ros 2006; Fan et al. 2006; Dai et al. 2006; Li 2007;
Ghisellini et al. 2006,2007; Wang et al. 2006; Gupta \& Zhang 2007; Murase et al. 2006;
Amati et al. 2006; Liang et al. 2007). Future detectors more sensitive to {\em Swift}
BAT, e.g. EXIST (Grindley 2006), would detect much more of these events and greatly
increase the sample of LL-GRBs.

\section{Conclusions}
We have derived the luminosity function and the local rate of the HL-GRBs with the {\em
Swift} $z$-known GRB sample and explored the possibility that LL-GRBs as a unique GRB
population from the HL-GRBs. We find that $\rho_0^{\rm LL}$ is $\sim 325_{-177}^{+352}$
Gpc$^{-3}$ yr$^{-1}$, which is much higher than $\rho_0^{\rm HL}$($1.12_{-0.20}^{+0.43}$
Gpc$^{-3}$ yr$^{-1}$). This rate is $\sim 0.7\%$ of the local Type Ib/c SNe. Our results,
together with the finding that less than 10\% of Type Ib/c SNe are associated with
off-beam GRBs, suggest that LL-GRBs have a beaming factor typically less than 14, or a
jet angle typically wider than 31$^{\rm o}$. The high local GRB rate, the small beaming
factor, and low luminosity make the LL-GRBs distinct from the HL-GRBs. Although the
current data could not fully rule out the possibility that both HL- and LL-GRBs are
the same population, our results suggest that LL-GRBs are likely a unique GRB population
and the observed low redshift GRB sample is dominated by LL-GRBs.

We note that Guetta \& Della Valle (2007) reproduce the three main results, i.e.,
the high local rate of LL-GRBs, HL- and LL-GRBs as two populations, and wide jet
opening angle for LL-GRBs, three months after we posted this paper to astro-ph.
They also suggest that the possibility of one population for both the HL- and
LL-GRBs (Guetta et al. 2004) cannot be convincingly ruled out.

\acknowledgments We thank Don Lamb for helpful discussion. This work was supported by
NASA under grants NNG05GB67G, NNG05GH92G and NNG05GH91G (EWL, BZ \& ZGD), and the
National Natural Science Foundation of China under grants 10463001 (EWL) and 10221001 \&
10233010 (ZGD).

\clearpage

\clearpage

\clearpage

\clearpage


\begin{thebibliography}{99}
\bibitem[]{1}Amati, L., Della Valle, M., Frontera, F. , Malesani, D.,  et al. 2006, A\&A, submitted (astro-ph/0607148).
\bibitem[]{2}Band, D.,Matteson, J., Ford, L., Schaefer, B.,  et al. 1993, ApJ, 413, 281
\bibitem[]{3}Band, D. L. 2003, ApJ, 588, 945
\bibitem[]{4}Bloom, J. S. 2003, AJ, 125, 2865
\bibitem[]{5}Bloom, J. S.,Kulkarni, S. R., Djorgovski, S. G., Eichelberger, A. C., et al. 1999, Nature, 401, 453
\bibitem[]{6}Bloom, J. S. et al. 2002, ApJ, 572, L45
\bibitem[]{7}Bloom, J. S., Frail, D. A., \& Sari, R. 2001, AJ, 121, 2879
\bibitem[]{8}Campana, S.,Mangano, V., Blustin, A. J., Brown, P., Burrows, D. N.,  et al. 2006, Nature, 442, 1008
\bibitem[]{9}Cappellaro, E., Evans, R., \& Turatto, M. 1999, A\&A, 351, 459
\bibitem[]{10}Cobb, B. E.,Bailyn, C. D., van Dokkum, P. G., \& Natarajan, P. 2006, ApJ, 645, L113
\bibitem[]{11}Colgate, S. A. 1974, ApJ, 187, 333
\bibitem[]{12}Dai, X. \&  Zhang, B. 2005, ApJ, 621, 875
\bibitem[]{13}Dai, Z. G, Zhang, B., \& Liang, E. W, 2006, astro-ph/0604510
\bibitem[]{14}Daigne, Fr\'{e}d\'{e}ric; Rossi, E. M., Mochkovitch, R., 2006, MNRAS, 372, 1034
\bibitem[]{16}Della Valle, M.,Malesani, D., Benetti, S., Testa, V., Hamuy, M., et al. 2003, A\&A, 406, L33
\bibitem[]{15}Della Valle, M.,Chincarini, G. , Panagia, N. , Tagliaferri, G., et al. 2006, Nature, 444,1050
\bibitem[]{17}Fan, Y. Z., Piran, T., \& Xu, D. 2006, JCAP, 9, 13.
\bibitem[]{19}Folkes, S.,Ronen, S., Price, I., Lahav, O., et al. 1999, MNRAS, 308, 459
\bibitem[]{20}Frail, D.,Kulkarni, S. R., Sari, R., Djorgovski, S. G., et al. 2001, ApJ, 562, L55
\bibitem[]{21}Fynbo, J. P. U., Watson, D., Thoene, C. C., Sollerman, J., et, al. 2006, Nature, 444, 1047
\bibitem[]{22}Fynbo, J. P. U., Sollerman, J., Hjorth, J., Grundahl, F., et al. 2004, ApJ, 609, 962
\bibitem[]{23}Galama, T. J., Briggs, M. S., Wijers, R. A. M., Vreeswijk, P. M., et al. 1998, Nature, 395, 670
\bibitem[]{24}Gal-Yam, A., Fox, D., Price, P., Davis, M., et al. 2006, Nature, 444, 1053
\bibitem[]{25}Gehrels, N., Norris, J. P., Mangano, V., Barthelmy, S. D., et al. 2006, Nature, 444, 1044
\bibitem[]{27}Ghisellini, G., Ghirlanda, G., Mereghetti, S., Bosnjak, Z., et al. 2006, MNRAS, 372, 1699
\bibitem[]{26}Ghisellini, G., Ghirlanda, G., Tavecchio, F. 2007, MNARS, 375, L36
\bibitem[]{28}Grindley, J. E. 2006, in S. S. Holt, N. Gehrels \& J. A. Nousek (eds), AIP Conf. Proc., 836, 631
\bibitem[]{29}Guetta, D., Perna, R., Stella, L.,  \& Vietri, M. 2004, ApJ, 615, L73
\bibitem[]{30}Guetta, D., Piran, T., \& Waxman, E. 2005, ApJ, 619, 412
\bibitem[]{301}Guetta, D. \& Della Valle, M., 2007, ApJ, in press (astro-ph/0612194).
\bibitem[]{31}Gupta, N. \& Zhang, B., Astroparticle Physics, 2007, in press(astro-ph/0606744)
\bibitem[]{32}Hjorth, J., Sollerman, J., Moller, P., Fynbo, J. P. U., et al. 2003, Nature, 423, 847
\bibitem[]{33}Ioka, K.,  Nakamura, T. 2001, ApJ, 554, L163
\bibitem[]{34}Jakobsson, P., Levan, A., Fynbo, J. P. U., Priddey, R, et al. 2006, A\&A, 447, 897
\bibitem[]{35}Kulkarni, S. R., Frail, D. A., Wieringa, M. H., Ekers, R. D.,  et al. 1998, Nature, 395, 663
\bibitem[]{36}Lamb, D. Q., Donaghy, T. Q. \& Graziani, C. 2005, ApJ, 620, 355
\bibitem[]{361}Le, T. \& Dermer, C., D. 2007, ApJ, in press (astro-ph/0610043)
\bibitem[]{37}Li, L. X., 2007, MNRAS, 375, 240
\bibitem[]{38}Liang, E. W., Zhang, B. B., Stamatikos, M., Zhang, B., et al. 2007, ApJ, 653, L81
\bibitem[]{39}Lloyd-Ronning N. M., Fryer, C., \& Ramirez-Ruiz E. 2002, ApJ, 574, 554
\bibitem[]{40}Lloyd-Ronning, N. M., Dai, X. Y., \& Zhang, B. 2004, ApJ, 601, 371
\bibitem[]{41}M\'esz\'aros, P. 2006, Rep. on Prog. in Phys. 69, 2259
\bibitem[]{42}Marzke, R. O., da Costa, L. N., Pellegrini, P. S., Willmer, C. N. A., \& Geller, M. J. 1998, ApJ, 503, 617
\bibitem[]{43}Mazzali, P. A., Deng, J. S., Nomoto, K., Sauer, D. N., et al. 2006, Nature, 442, 1018
\bibitem[]{431}Malesani, D., Tagliaferri, G., Chincarini, G. Covino, S. et al. 2004, ApJ, 609, L5
\bibitem[]{44}Mirabal, N., Halpern, J. P., An, D., Thorstensen, J. R., \&  Terndrup, D. M. 2006, ApJ, 643, L99
\bibitem[]{45}Modjaz, M., Stanek, K. Z., Garnavich, P. M., Berlind, P., et al. 2006, ApJ, 645, L21
\bibitem[]{46}Murase, K. Ioka, K., Nagataki, S., Nakamura, T. 2006, ApJ, 651, L5
\bibitem[]{47}Nakamura, T. 1999, ApJ, 522, L101
\bibitem[]{48}Norris, J. P.,Bonnell, J. T., Kazanas, D., Scargle, J. D.,  et al. 2005, ApJ, 627, 324
\bibitem[]{50}Norris, J.P. 2002, ApJ, 579, 386
\bibitem[]{51}Pian, E., Mazzali, P. A., Masetti, N., Ferrero, P., et al. 2006, Nature, 442, 1011
\bibitem[]{52}Piran, T. 2005, Rev. Mod. Phy., 76, 1143
\bibitem[]{53}Porciani,  C. \& Madau, P. 2001, ApJ, 548, 522
\bibitem[]{54}Press, W. H., et al 1999, Numerical Recipes in Fortran, Cambridge University Press
\bibitem[]{56}Sakamoto, T., et al. 2006, ApJ, submitted
\bibitem[]{57}Sazonov, S. Y., Lutovinov, A. A., Sunyaev, R. A., 2004, Nature, 430, 646
\bibitem[]{58}Schmidt, M. 2001, ApJ, 552, 36
\bibitem[]{59}Soderberg, A. M., et al. 2002, GCN Circ. 1554
\bibitem[]{60}Soderberg, A. M., Frail, D. A., \& Wieringa, M. H. 2004a, ApJ, 607, L13
\bibitem[]{61}Soderberg, A. M., Kulkarni, S. R., Berger, E., Fox, D. W., et al. 2004b, Nature, 430, 648
\bibitem[]{63}Soderberg, A. M., Nakar, E., Berger, E., Kulkarni, S. R. 2006a, ApJ, 638, 930
\bibitem[]{64}Soderberg, A. M., Kulkarni, S. R. Nakar, E., Berger, E., et al. 2006b, Nature, 442, 1014
\bibitem[]{65}Sollerman, J., Jaunsen, A. O., Fynbo, J. P. U., Hjorth, J., et al. 2006, A\&A, 454, 503
\bibitem[]{66}Stanek, K. Z., Matheson, T., Garnavich, P. M., Martini, P, et al. 2003, ApJ, 591, L17
\bibitem[]{67}Stern, B. E., Atteia J.-L., \& Hurley K. 2002, ApJ, 578, 304
\bibitem[]{68}Stern, B. E., Poutanen, J., \& Svensson, R. 1999, ApJ, 510, 312
\bibitem[]{69}Tinney, C. 1998, IAUC 6896, 1
\bibitem[]{70}Toma, K., Ioka, K., Sakamoto, T., Nakamura, T. 2007, ApJ, in press(astro-ph/0610867)
\bibitem[]{71}Wang, X. Y \& M\'{e}sz\'{a}ros, P., 2006, ApJ, 643, L95
\bibitem[]{72}Wang, X. Y., Dai, Z. G., Lu, T., Wei, D. M., \& Huang, Y. F. 2000, A\&A, 357, 543
\bibitem[]{73}Wang, X. Y., Li, Z., Waxman, E. \& M\'esz\'aros, P. 2006, ApJ, submitted(astro-ph/0608033)
\bibitem[]{74}Waxman, E. 2004a, ApJ, 602, 886
\bibitem[]{75}Waxman, E. 2004b, ApJ, 605, L97
\bibitem[]{76}Woosley, S. E. \& Bloom, J. 2006, ARA\&A, 44, 507
\bibitem[]{77}Woosley, S. E. 1993, ApJ, 405, 273
\bibitem[]{78}Yamazaki, R., Yonetoku, D., \& Nakamura, T. 2003, ApJ, 594, L79
\bibitem[]{79}Zeh, A., Klose, S., Hartmann, D. H., et al. 2004, ApJ, 609, 952
\bibitem[]{} Zhang, B. 2006, Nature, 444, 1010
\bibitem[]{80}Zhang, B., Zhang, B. B., Liang, E. W. , Gehrels, N., Burrows, D. N. \& M\'esz\'aros, P.
2007a, ApJ, 655, L25
\bibitem[]{801} Zhang, B., Liang, E. W., Page, K. L., Grupe, D. et al. 2007b, ApJ, 655, 989
\bibitem[]{81}Zhang, B., \& M\'{e}sz\'{a}ros, P., 2004, Int. J. Mod. Phy. A, 19, 2385
\bibitem[]{82}Zhang, B., Dai, X., Lloyd-Ronning, N. M. \& M\'esz\'aros, P. 2004, ApJ, 601, L119
\bibitem[]{84}Zhang, W., Woolsey, S. E. \& MacFadyen, A. I. 2003, ApJ, 586, 356
\end{thebibliography}
\end{document}